\documentclass[aps,numbers,amsmath,amssymb,nofootinbib,superscriptaddress,floatfix,showkeys,twocolumn,longbibliography]{revtex4-1}
\usepackage{booktabs}
\usepackage{float}
\usepackage{threeparttable}
\usepackage{graphicx}% Include figure files
\usepackage{dcolumn}% Align table columns on decimal point
\usepackage{bm}% bold math
\usepackage{captcont}
\usepackage{xcolor}
\usepackage{threeparttable}

\usepackage{epstopdf}
\usepackage{mathtools}
\usepackage{natbib}
\usepackage{ulem}
\usepackage{color,txfonts}
\definecolor{blue0}{rgb}{0,0,0.6}
\usepackage[colorlinks,linkcolor=blue0,anchorcolor=blue0,citecolor=blue0,urlcolor=blue0]{hyperref}

\usepackage{color,amsmath,amssymb,graphicx,latexsym,lineno}
\usepackage{threeparttable,multirow,txfonts,subfigure,booktabs}

\newcommand{\eV}{\mathrm{eV}}
\newcommand{\keV}{\mathrm{keV}}

\newcommand*{\mkred}[1]{{\color{blue}{#1}}}

\begin{document}
\title{%Searching Axion Through X-rays from Alpha Centauri
%Stringent Constraints on Axion-Like Particles with X-ray Observations of Alpha Centauri
Searching for Axion-Like Particles with X-ray Observations of Alpha Centauri}

\author{Yu-Xuan Chen}
%$^{a,b}$
\author{Lei Lei}
\affiliation{Key Laboratory of Dark Matter and Space Astronomy, 
Purple Mountain Observatory, Chinese Academy of Sciences, Nanjing 210033, China}
\affiliation{School of Astronomy and Space Science, University of Science and Technology of China, Hefei, Anhui 230026, China}

\author{Zi-Qing Xia}

\author{Ziwei Wang}
\email{Corresponding author. zwwang@pmo.ac.cn}
\affiliation{Key Laboratory of Dark Matter and Space Astronomy, 
Purple Mountain Observatory, Chinese Academy of Sciences, Nanjing 210033, China}

\author{Yue-Lin Sming Tsai}
\email{Corresponding author. smingtsai@pmo.ac.cn}
\author{Yi-Zhong Fan}
\email{Corresponding author. yzfan@pmo.ac.cn}

\affiliation{Key Laboratory of Dark Matter and Space Astronomy, 
Purple Mountain Observatory, Chinese Academy of Sciences, Nanjing 210033, China}
\affiliation{School of Astronomy and Space Science, University of Science and Technology of China, Hefei, Anhui 230026, China}

\begin{abstract}
We investigate the production of axion-like particles (ALPs) in stellar cores, where they interact with electromagnetic fields and electrons, with typical masses between $\mathcal O(0.1)$ and $\mathcal O(10)$ keV. These low-energy ALPs are gravitationally trapped in the orbits of stars and subsequently decay into two photons that we detect as monochromatic X-ray lines. We propose to search for these gravitationally trapped ALPs in the Alpha Centauri binary system, our closest stellar neighbor, using sensitive X-ray detectors like Chandra and eROSITA. Our search for ALP decay signals in the energy range of $0.2$ keV to $10$ keV yielded null results, thus establishing the most stringent limits on ALP interactions to date. 
%Specifically, if ALPs are mainly produced by Compton or bremsstrahlung processes (ALP-electron coupling $g_{aee}$ being significant), we have improved the limits on the ALP-photon coupling $g_{a\gamma\gamma}$ by \mkblue{one to two} orders of magnitude, in ALP mass range between $0.2~\keV$ to $5~\keV$, compared to previous measurements, including those from GW170817, SN 2023ixf, and other sources.
In the case of ALP-electron coupling $g_{aee}\leq 10^{-15}$, 
we have improved the limits on the ALP-photon coupling $g_{a\gamma\gamma}$ in ALP mass range between $0.25~\keV$ and $5~\keV$, compared to previous measurements, including those from GW170817, SN 2023ixf, and other sources, and specially the improvement reaches about 2 orders of magnitude at the mass of 2 keV. Even tighter constraints are set for larger $g_{aee}$.
\end{abstract}

\date{\today}

\maketitle

%\section
{\it Introduction.} 
Axion-like particles (ALPs) are hypothetical particles that possess similar properties to axions~\cite{Weinberg:1977ma,Wilczek:1977pj,Dine:1981rt,Reece:2023czb}. They are well motivated and arise in fundamental theories, including string theory~\cite{Svrcek:2006yi,Arvanitaki:2009fg}.
As pseudoscalar particles, ALPs can couple to photons or other Standard Model charge particles with a wide allowed range of masses and coupling strengths~\cite{Graham:2015ouw,Irastorza:2018dyq,Choi:2020rgn}.
Recently, ALPs have attracted significant attention, leading to extensive efforts to explore them through ground 
%increasing attention, and great efforts have been made to probe ALPs through both ground
experiments~\cite{Redondo:2010dp,Ehret:2010mh,Bahre:2013ywa,IAXO:2019mpb,IAXO:2020wwp,Isleif:2022ytq} and astrophysical observations~\cite{Hooper:2007bq,Fermi-LAT:2016nkz,Xia:2018xbt,Liang:2018mqm,Majumdar:2018sbv,Chen:2019fsq,Meyer:2020vzy,Yuan:2020xui,Li:2020pcn}.

The interaction of ALPs $a$ with electromagnetic fields $F$ (its dual marked as $\tilde F$) and electrons $\psi$ can be described by the Lagrangian
\begin{equation}
    {\mathcal L} = \frac{1}{2}\partial^\mu a\partial_\mu a - \frac{1}{2} m_a^2a^2 - \frac{1}{4} g_{a\gamma\gamma}aF^{\mu\nu}\widetilde{F}_{\mu\nu} - \frac{1}{2}g_{aee}a\bar\psi\gamma^{5}\psi
\,
\end{equation}  
where $m_a$ is the mass of ALPs, $g_{a\gamma\gamma}$  and $g_{aee}$ are the coupling constant of ALPs to photons and electrons separately.
ALPs are efficiently produced in stellar plasmas and can escape from stars due to their extremely weak interaction with matter. These particles are generated over a broad mass range with keV-level energies, depending on the stellar core temperature. Stellar ALPs may eventually decay or convert into X-rays, giving a promising signal for detecting axions and ALPs. The CERN Axion Solar Telescope (CAST) is one of the most powerful experiments for detecting axions and ALPs produced in the Sun~\cite{CAST:2017uph}. This method has also been applied to study super star clusters~\cite{Dessert:2020lil}, helium-burning stars in globular clusters~\cite{Lucente:2022wai}, Betelgeuse~\cite{Xiao:2020pra,Xiao:2022rxk}, supernovae (SNe)~\cite{Payez:2014xsa,Caputo:2022mah,Hoof:2022xbe,Muller:2023pip,Caputo:2021rux, Diamond:2023scc} and neutron star (NS) merger~\cite{Dev:2023hax, Diamond:2023cto} to constrain the ALP-photon coupling. 
In supernovae, ALPs can be produced with MeV energies and subsequently decay or convert into gamma rays.

%\begin{figure}
%\centering 
%\includegraphics[width=1.0\linewidth]{mkFigs/ALP_sketch_new.png}
%\caption{A schematic diagram illustrating stellar ALPs gravitationally trapped by the Alpha Centauri system. For produced ALPs with a mass of $m_a=1$~keV, the number density per unit time and per unit energy as a function of ALP kinetic energy $T_a$ is depicted in the lower right corner. The brown region represents the gravitationally trapped ALPs, whose velocities fall below the escape threshold.}
%\label{fig:schematic} 
%\end{figure}

For stellar ALPs with masses between $\mathcal O(0.1)$ and $\mathcal O(10)$ keV, low-energy final states with velocities below the escape threshold are gravitationally trapped in the orbit of star. %\sout{\textcolor{red}{, as schematically illustrated in Fig. \ref{fig:schematic}}}.
The number density of ALPs increases over the star lifetime, enhancing ALP-induced X-ray emissions. This has prompted by some groups to constrain the properties of stellar ALPs using solar observations from SphinX and NuSTAR \cite{VanTilburg:2020jvl, DeRocco:2022jyq, Beaufort:2023zuj}. However, due to the brightness of the Sun, the emissions cannot be directly observed by Chandra or other sensitive X-ray telescopes, limiting solar observations to the relatively massive ALPs with greatly reduced sensitivity.

This work aims to {search ALP signals} within the keV mass range by focusing on precise X-ray measurements of distant stars with temperatures and lifetimes comparable to the Sun. Higher temperatures correlate with stronger ALPs radiation, while longer lifetimes allow for greater ALPs accumulation. Thus, the Alpha Centauri binary system, our closest stellar neighbors consisting of Alpha Centauri A and B \cite{1999JKAS...32..119K}, emerges as an ideal target. Its greater distance enables effective detection by sensitive instruments on Chandra and eROSITA \cite{Torres:2006tk, 2017A&A...598L...7K, 2020ApJS..250...16A, Merloni:2024zgn, eROSITA:2020emt}. Although Sirius A has some favorable aspects, it is excluded from our investigation due to its relatively young age of approximately 240 million years, which limits the number of accumulated ALPs. As summarized in Table \ref{Table 1}, other nearby stars are too cold to serve as suitable targets.

\begin{table}
    \centering
    \renewcommand{\arraystretch}{1.2}
    \setlength{\tabcolsep}{6pt}
    \caption{The basic information of some nearby stars.}
    \label{Table 1}
    \begin{threeparttable}
        \begin{tabular}{c|c c c c } 
            \hline\hline
            Name\tnote{\dag} & ${\rm D}$\tnote{$\star$} \, [pc] & ${\rm M}$\tnote{$\clubsuit$} \, [${\rm M_{\odot}}$] & ${\rm T}$\tnote{$\divideontimes$} \, [K] & Age \, [Gyr] \\ 
            \hline
            Sun & 4.85$\times10^{-6}$ & 1.00 & 5772 & 4.60\\ 
            Proxima Centauri & 1.30 & 0.12 & 2992 & 4.85\\ 
            Alpha Centauri A & 1.33 & 1.08 & 5804 & 5.26\\ 
            Alpha Centauri B & 1.33 & 0.91 & 5207 & 5.26\\ 
            Barnard's Star & 1.83 & 0.16 & 3195 & $\sim$ 10\\ 
            Wolf 359 & 2.41 & 0.11 & 2749 & 0.1$-$1.5\\
            Lalande 21185 & 2.55 & 0.39 & 3547 & 8.05\\ 
            Sirius A & 2.64 & 2.06 & 9845 & 0.24\\ 
            \hline\hline
        \end{tabular}
    \begin{tablenotes}
        \footnotesize
        \item[\dag] The stars we list have masses greater than 0.1 ${\rm M_{\odot}}$ and the furthest distance reaches up to Sirius A. The basic information of these nearest stars can be found in  %\url{https://en.wikipedia.org/wiki/List_of_nearest_stars} 
        \mkred{Ref.~\cite{2015arXiv151007674M,2002A&A...390.1115B,Bobylev:2010tg,2018A&A...616A..37B,Bobylev:2022,2021AJ....162...14A,2018ApJ...864...99J,2017A&A...598L...7K,2021ApJ...918...40P,2024A&A...690A..79G,2017ApJ...840...70B,2011PASA...28...58D}.}
        \item[$\star$] ${\rm D}$ is the distance from the Earth to the star.
        \item[$\clubsuit$] ${\rm M}$ is the star's mass.
        \item[$\divideontimes$] ${\rm T}$ is the star's effective temperature.
        
      \end{tablenotes}
    \end{threeparttable}

\end{table}
 
We carry out the spectral line analysis to search for the stellar ALP-induced signal, and report the null result  
in the mass range $\mathcal{O}(0.1) - \mathcal{O}(10)$ keV. 
In addition to Ref.~\cite{DeRocco:2022jyq, Beaufort:2023zuj}, 
we also consider bremsstrahlung radiation, which may dominate the signal at energies below approximately $2$ keV, significantly improving the constraints in this range.
Our findings set the most stringent limits on the ALP-photon coupling in this ALP mass range, regardless of dominant ALP production processes. 
%In particular, limits on the coupling in the energy range of $0.2-5$ keV have been improved by at least {two to three} orders of magnitudes compared to those reported in Ref.~\cite{Beaufort:2023zuj} when the ALP-electron coupling is large enough that the dominant ALP production process is either Compton or bremsstrahlung radiation.
In the case of ALP-electron coupling $g_{aee}\leq 10^{-15}$, we have improved the limits by up to 2 orders of magnitude on the ALP-photon coupling $g_{a\gamma\gamma}$ in ALP mass range between $0.25~\keV$ to $5~\keV$, compared to previous limits from Horizontal Branch (HB) stars in globular clusters~\cite{Carenza:2020zil}, SN 2023ixf~\cite{Muller:2023pip}, GW170817~\cite{Dev:2023hax} and our sun~\cite{Beaufort:2023zuj}. For larger $g_{aee}$ the constraints are even tighter.

%%%%%%%%%%%%%%%%%%%%%%%%%%%%%%%%%%%%%%%%%%%%%%%%%%%%%%%%%%%%%%%%%%%%%%%%%%%%%%%%%
%\section{Stellar ALP induced X-ray from Alpha Centauri}
%\section
{\it ALP-induced X-rays from Alpha Centauri.}
%%%%%%%%%%%%%%%%%%%%%%%%%%%%%%%%%%%%%%%%%%%%%%%%%%%%%%%%%%%%%%%%%%%%%%%%%%%%%%%%%
Within the Alpha Centauri A and B, a large amount of ALPs can be produced in the stellar plasma through the Primakoff effect, photon coalescence, Compton mechanism, 
and bremsstrahlung radiation.
The Primakoff effect works when a photon transforms into an ALP in a strong electric or magnetic field, as in the stellar plasmas of Alpha Centauri ~\cite{Raffelt:1985nk,DiLella:2000dn,Bastero-Gil:2021oky}. 
Moreover, photon coalescence $\gamma \gamma \rightarrow a$~\cite{Cadamuro:2010cz,Cadamuro:2011fd} in the dense stellar core can also result in rich ALP production,  
depending on star temperature, ALP mass, and $g_{a \gamma \gamma}$. 
Finally, once the coupling $g_{aee}$ is considered,  
the Compton and bremsstrahlung mechanisms also trigger ALP productions~\cite{Zhitnitsky:1980tq,Dine:1981rt,Beaufort:2023zuj}. 
The total ALP production rate $\Gamma_{\rm Prod.}$ is the sum of the transition rates from the Primakoff process $\Gamma^{\rm Primakoff}$, photon coalescence $\Gamma^{\rm coal.}$, the Compton mechanism $\Gamma^{\rm Comp.}$ and the bremsstrahlung radiation $\Gamma^{\text{brem.}}$,
\begin{equation}
\Gamma_{\rm Prod.}=\Gamma^{\rm Primakoff}+\Gamma^{\rm coal.}+\Gamma^{\rm Comp.} + \Gamma^{\text{brem.}},
\label{Gamma_prod}
\end{equation}
with detailed expressions available in the Supplemental Material.%~\cite{SupplementalMaterial}. 

\begin{figure}[!ht]
%\centering 
\includegraphics[width=0.55\textwidth]{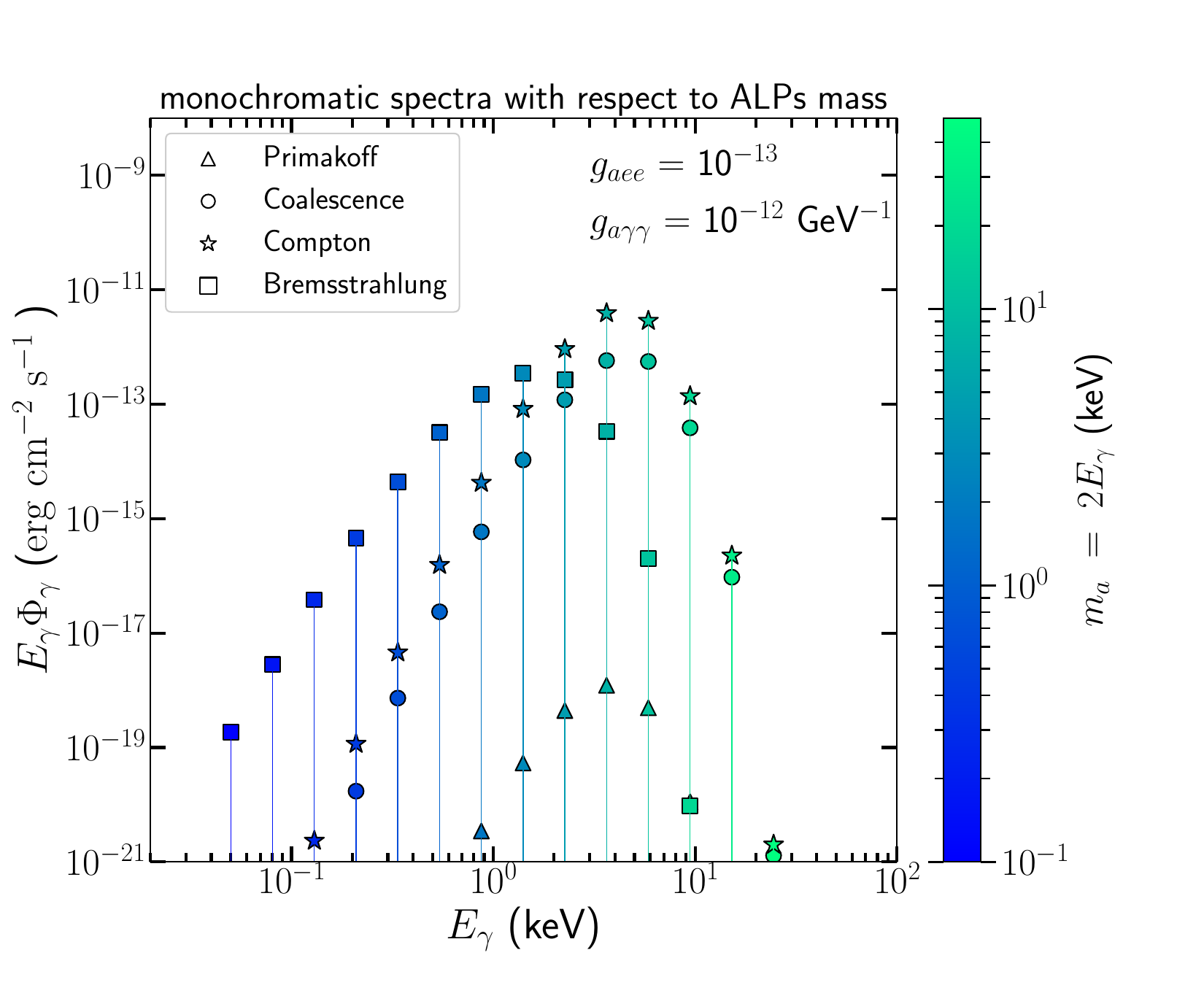}
\caption{The line spectra of ALP-induced photons as a function of photon energy for Alpha Centauri binary, 
with different masses indicated by the color bar. 
The spectra result from four components: the Primakoff effect (triangles), photon coalescence (circles), the Compton mechanism (pentagons) and bremstrahlung (squares). 
The ALP-electron and ALP-photon couplings are set to $g_{aee}=10^{-13}$ and $g_{a\gamma\gamma}=10^{-12}~{\rm GeV}^{-1}$. 
%The 95\% C.L. flux upper limits for Alpha Centauri A (Chandra data) and the Alpha Centauri binary (eROSITA data) are represented by the red and orange solid lines, \mkred{respectively with the method provided in the next section}. 
Notice that varying $g_{aee}$ affects the Compton and bremsstrahlung fluxes; when $g_{aee}$ is significantly small, photon coalescence can become dominant.}
\label{Fig:photon_spectrum} 
\end{figure}

ALPs produced in the star span a large velocity phase space. For ALPs with large momentum, they escape from the gravitational potential. 
%(A spatial smeared X-ray background at the scale larger than cluster is predicted from decaying of high energy ALPs). 
Despite the suppressed efficiency of ALPs production rate,  
%with velocities smaller than the escaping threshold, 
non-relativistic ALPs can be captured by the gravity of Alpha Centauri, where they accumulate in orbit over billions of years, significantly increasing their density to  
\begin{equation}
n_a^{\rm orb}(t)=\frac{dn_a^{\rm trap}}{dt}\times \frac{1-\exp\left(-\Gamma_{a \gamma \gamma}t\right)}{\Gamma_{a \gamma \gamma}},  
\label{axions_in_orbit_ori}
\end{equation}
where $\Gamma_{a \gamma \gamma}$ is denoted as the decay rate of ALPs into photons, with units of ${\rm s}^{-1}$.    
The number density of ALPs trapped in the gravitational system per unit time $t$ is presented as $dn_a^{\rm trap}/dt$.
%as illustrated with $m_a=1~\keV$ in the inset plot of Fig.~\ref{fig:schematic}. 
These quantities are evaluated in Supplemental Material. 
We note that the density of ALPs cloud trapped around the star takes the profile of $r^{-4}$ with the distance to the star $r$, 
thus ALPs produced from Alpha Centuri A and B mostly locates nearby. 
Their orbits are gravitationally stable, allowing trapped ALPs to accumulate over billions of years. We treat Alpha Centauri as a point source due to its far distance from Earth, then the ALP-induced photon flux can be written as:
\begin{equation}
    \Phi_{E,\gamma}(t) = \frac{1}{4\pi D^2} \int_{R_s}^{\infty} d{r}~4\pi{r}^2 n_a^{\rm orb}(t) \Gamma_{a\gamma\gamma},
\label{eq:acen_flux}
\end{equation} 
where $R_{s}$, $r$ and $D$ are radius of Alpha Centuari A or B, the distance between trapped ALP clouds and stellar core, and the distance to Earth, respectively.

We use the \texttt{MESA} code~\cite{Paxton:2010ji,Paxton:2013pj} to calculate the radial profiles of temperature, density, and composition of Alpha Centauri stars as they evolve, in order to determine the ALP-induced photon spectrum. 
For Alpha Centauri A, the initial metallicity is $Z_A = 0.027$ and mass $M_A = 1.079\,M_{\odot}$~\cite{2018ApJ...864...99J,2021AJ....162...14A}; 
for Alpha Centauri B, the initial values are $Z_B = 0.027$ and $M_B = 0.909\,M_{\odot}$~\cite{2018ApJ...864...99J,2021AJ....162...14A}.
We track their evolution from the begining of main sequence to the present, neglecting the whole pre-main sequence phase of the stars, since production rate of ALPs are highly suppressed at that time due to cold temperature of the stellar core.

%%%%%%%%%%%%%%%%%%%%%%%%%%%%%%%%%%%%%%
\begin{figure*}[!ht]
\centering 
\includegraphics[width=0.99\linewidth]{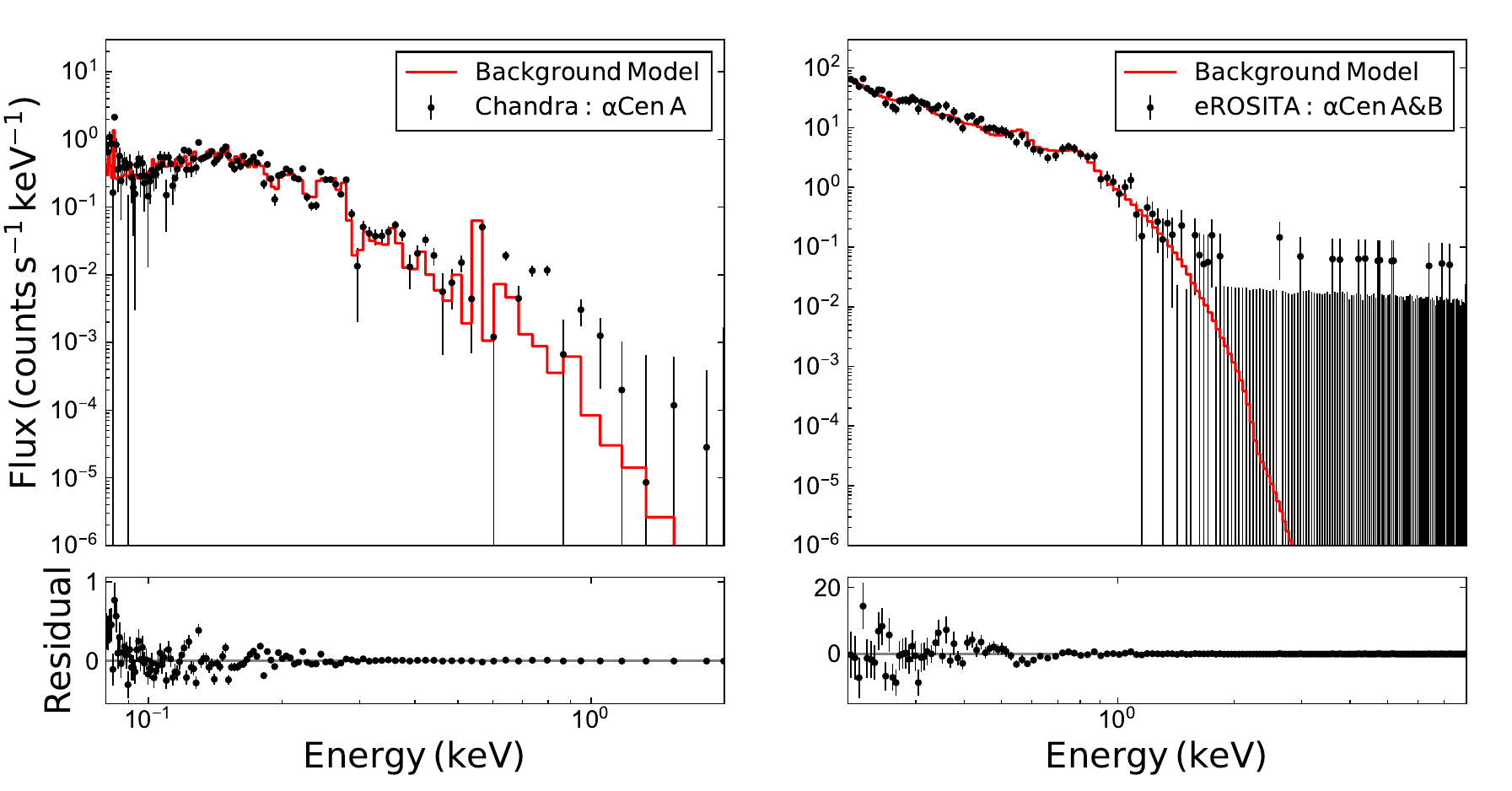}
\caption{The spectral data (black error-bars) and the best-fit background models (red solid lines). Adaptive rebinning of the X-ray photon counts from Chandra (eROSITA) X-ray satellite is adopted to balance the signal-to-noise ratio and the bin width.
\textbf{Upper Left panel:} Chandra X-ray spectrum of Alpha Cen A.
\textbf{Upper Right panel:} eROSITA X-ray spectrum of Alpha Cen A and B.  
%We perform adaptive rebinning of the X-ray photon counts from Chandra (eROSITA) X-ray satellite by successively grouping energy bins until a signal-to-noise ratio of 100 (7) is achieved. To prevent excessive bin width, groupings are capped at a maximum of 100 (7) original bins per final bin.
\textbf{Bottom panels:} The residual of the modeling of two datasets.
}
%Bins in plot are plotted that have at least signal-to-noise ratio $\geq 7$ or are grouped in sets of 7 bins. }
\label{fig:spec} 
\end{figure*}

The theoretical uncertainty of the photon flux is assumed to be determined by stellar models, including variations in stellar mass, metallicity, age, convection models, and elemental diffusion. 
The maximum difference between the optimistic and conservative results reaches 55\% within the photon energy range from 0.08 to 8.0 keV, 
corresponding to a $\pm 14\%$ bias in $g_{a\gamma\gamma}$. 
A detailed comparison is provided in the Supplemental Material.

Fig.~\ref{Fig:photon_spectrum} illustrates the ALP-induced line spectra for various masses (represented by the color bar) from Alpha Centauri binary. 
The spectra appear as monochromatic lines due to decays from non-relativistic ALPs. 
With couplings set to $g_{a\gamma\gamma} = 10^{-12}~\text{GeV}^{-1} $ and $ g_{aee} = 10^{-13}$, the dominant contribution comes from the Compton mechanism for ALP masses above 4 keV and from bremsstrahlung for masses below 4 keV. As $g_{aee}$ is continuously reduced, both the Compton mechanism and bremsstrahlung are suppressed, until photon coalescence becomes dominant. 
The Primakoff process contributes significantly less to the total photon flux compared to the other mechanisms, because the transition rate through Primakoff process depends on the quadratic of the ALP momentum.
The trapped ALPs are non-relativistic, with extremely small momentum ${\mathcal{O}(1)}~\eV$ for $m_a=1~\keV$, which significantly suppresses the Primakoff process. Thus, ALPs produced by the Primakoff process can be negligible compared to photon coalescence, Compton mechanism and bremsstrahlung.

%%%%%%%%%%%%%%%%%%%%%%%%%%%%%%%%%%%%%%%%%%%%%%%%%%%%%%%%%%%%%%%%
%\section
{\it X-ray observations.}
%\label{x-ray}
%%%%%%%%%%%%%%%%%%%%%%%%%%%%%%%%%%%%%%%%%%%%%%%%%%%%%%%%%%%%%%%%
We utilize publicly available X-ray data of Alpha Centauri from the Chandra X-ray Observatory~\cite{2020ApJS..250...16A} and the eROSITA telescope array aboard the Spektrum Roentgen Gamma (SRG) satellite~\cite{2024A&A...682A..35T,Merloni:2024zgn,eROSITA:2020emt}. %~\footnote{Chandra: \url{https://cda.harvard.edu/chaser/mainEntry}} and the SRG/eROSITA satellite.%~\footnote{eROSITA: \url{https://erosita.mpe.mpg.de/dr1/erodat/catalogue/source_details/DR1_Main/em01_220150_020_ML00001_002_c010}}. 
Chandra, known for its high sensitivity, observed the Alpha Centauri spectrum in 1999, 2007, and 2011 using the Low Energy Transmission Grating (LETG). 
To optimize statistic strength, we only adopt Chandra observations in a range of 0.08-2.0 keV of Alpha Centauri A (\texttt{Obs ID: 29, PI: Brinkman; Obs IDs: 7432 \& 12332, PI: Ayres}), 
because Chandra can resolve the Alpha Centauri A and B spatially as well as the Alpha Centauri A dominates the total X-ray flux in the binary system. 
However, eROSITA cannot distinguish between Alpha Centauri A and B due to its lower spatial resolution, 
thus we apply the full binary data with an all-sky survey in the 0.2–8~keV soft X-ray band~\cite{2024A&A...682A..35T}. 
We combine exposure time for  each telescope, amounting to 80 kiloseconds for Chandra and 258 seconds for eROSITA , respectively. We adopt the public code \texttt{CIAO} \cite{CIAO} to get a spectrum of Alpha Centauri A star with the standard calibration and data reduction procedures. Additionally, we utilize eROSITA spectra of Alpha Centauri A and B from DR1, whose calibration has been done by eROSITA DR1 team.

Stars emit faint X-rays compared to their optical or infrared light, 
with these X-rays reflecting coronal activity linked to stellar spot cycles~\cite{2012A&A...543A..84R,Robrade:2016iup,2016Natur.535..526W,2017MNRAS.464.3281W}.
X-ray observations indicate that the average coronal plasma temperature of Alpha Centauri B is about 1.5-2.0 million Kelvin~\cite{2012A&A...543A..84R}. 
The stellar X-ray spectrum can be modeled using collisionally ionized diffuse gas models~\cite{2012A&A...543A..84R}, incorporating parameters like plasma temperature and the abundances of trace elements (e.g., C, N, O, Ne, Mg, Al, Si, S, Ar, Ca, Fe, Ni). 
We use the \texttt{Xspec} code~\cite{1996ASPC..101...17A} to fit known stellar spectral lines to calculate the statistical strength of our predicted spectra for Alpha Centauri. 

\begin{figure*}[!ht]
\centering 
\includegraphics[width=0.49\textwidth]{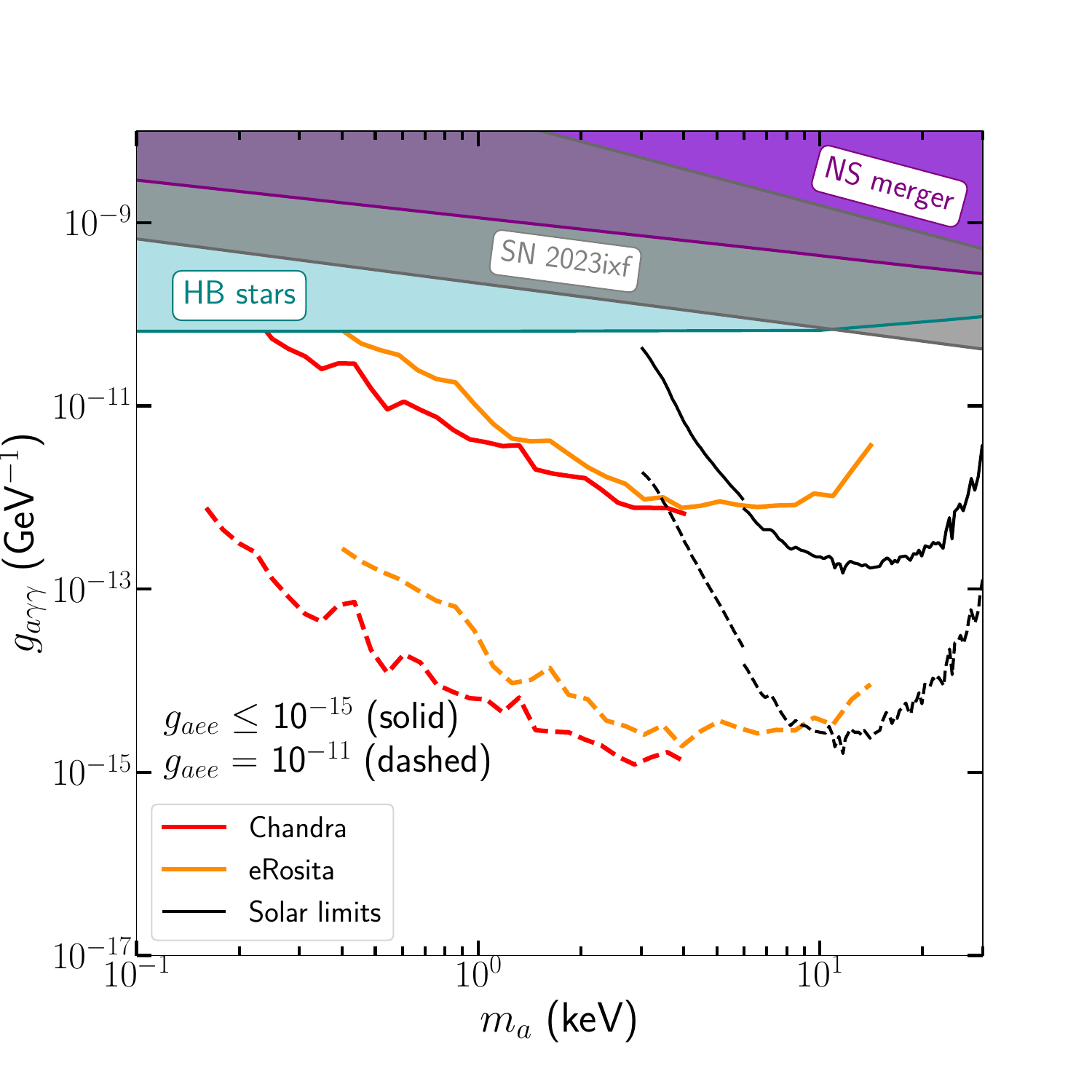}
\includegraphics[width=0.49\textwidth]{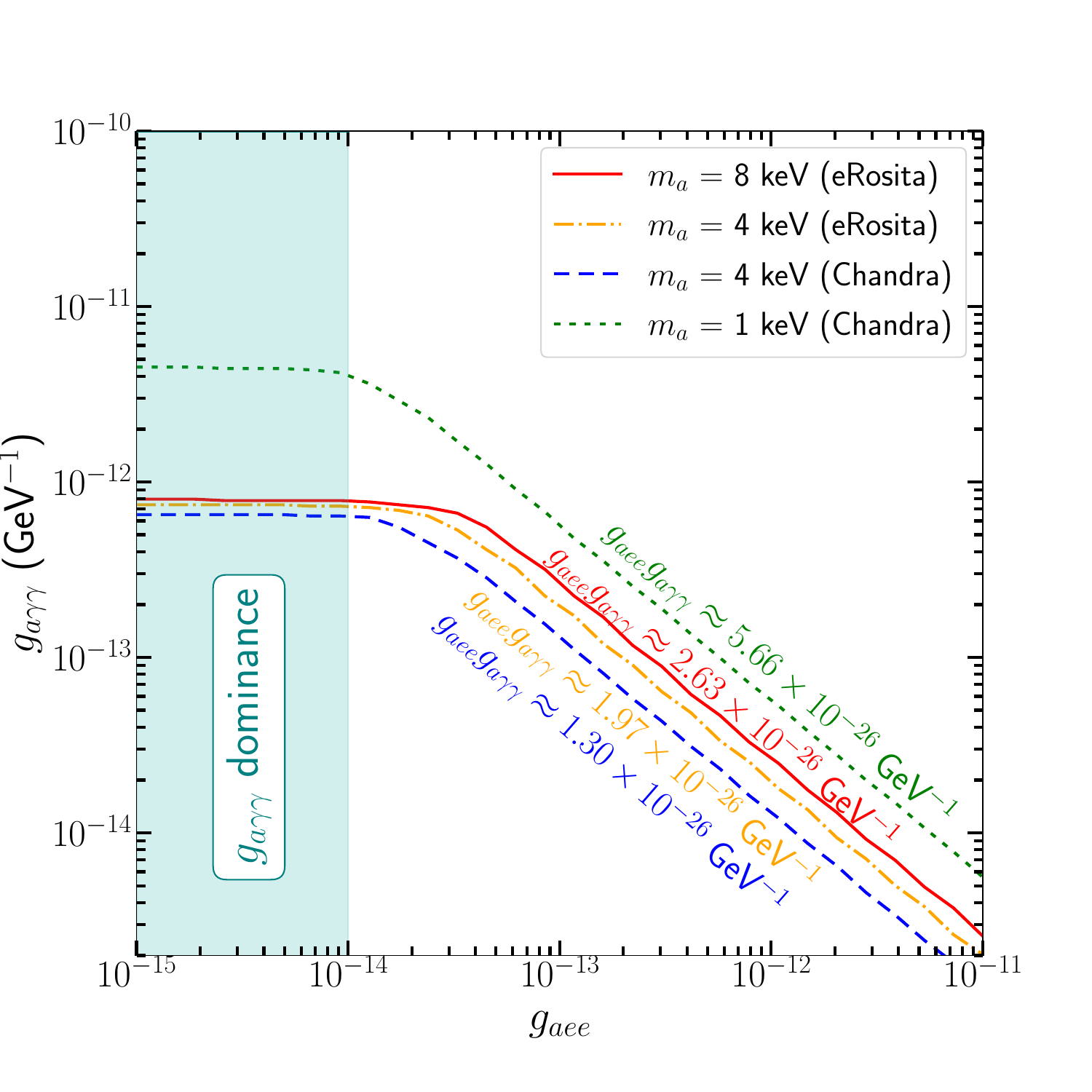}\\
\caption{Left panel: the ALP-photon coupling $g_{a \gamma \gamma}$ limits as a function of ALP mass $m_a$. The $95\%$ exclusion of
Chandra (red lines) and eROSITA (orange lines) are compared with the Solar X-ray limits (black lines)~\cite{Beaufort:2023zuj}. 
The solid and dashed lines refer to the ALP-electron coupling $g_{aee}\leq10^{-15}$ (namely, $g_{aee}$ can be ignored) and $g_{aee}=10^{-11}$, respectively.
The cyan, gray and purple regions corresponds to $95\%$ C.L. limit from HB stars in globular cluster~\cite{Carenza:2020zil}, Type II supernova SN 2023ixf~\cite{Muller:2023pip} and neutron star (NS) merger GW170817~\cite{Dev:2023hax}, respectively. 
Right panel: ALP-photon coupling $g_{a\gamma \gamma}$ limits as a function of ALP-electron coupling $g_{aee}$ for different ALP mass $m_a$, from Alpha Centauri X-ray measurements of Chandra and eROSITA. 
For eROSITA X-ray measurements, the ALP masses are $m_a = 4~{\rm keV}$ (orange dash-dotted line) and {$m_a = 8~{\rm keV}$} (red solid line), 
while for Chandra X-ray measurements, the ALP masses are $m_a = 1~{\rm keV}$ (green short-dashed line) and $m_a = 4~{\rm keV}$ (blue long-dashed line). The cyan region represents the photon coalescence-dominated ALP production.
}
\label{Fig:Cons} 
\end{figure*}

We analyze the LETG spectrum of Chandra in the 0.08–2.0 keV range using the BlackBody and Apec model \cite{1996ASPC..101...17A,2024NatAs...8..596K} in \texttt{Xspec} as the background model. 
The best-fit background model shows a plasma temperature of 0.13 keV with a maximum likelihood-based statistic $\mathcal{C}= 17730.52$ for 11902 bins. The statistic $\mathcal{C}$ is defined by \cite{1979ApJ...228..939C} for Poisson distributed data. 
%reduced chi-squared value of $\chi^2/\rm{d.o.f} = 0.89$, assuming solar abundances for the track elements. 
For eROSITA data on Alpha Centauri, we obtain the best-fit background model with $\mathcal{C}= 374.31$ for 776 bins in the 0.2–8.0 keV range. 
The comparison between the observed X-ray spectra and the best-fit background models are presented in Fig.~\ref{fig:spec} and 
more details are given in the Supplemental Materials.
Based on these background model parameters, we add Gaussian emission lines at energies $E_l$ to estimate the 95\% upper limit of the line flux. 
For each energy $E_l$, we calculate $\mathcal{C}$ by varying the line flux, using the energy resolution at $E_l$ as the Gaussian width. 

%%%%%%%%%%%%%%%%%%%%%%%%%%%%%%%%%%%%%%%%%%%%%%%%%%%%
%\section
{\it Result and Discussion.}
%%%%%%%%%%%%%%%%%%%%%%%%%%%%%%%%%%%%%%%%%%%%%%%%%%%%
After obtaining the best-fit background model for the observed spectra of Alpha Centauri stars, 
we search for possible emission lines within the energy range of 0.08–8.0 keV by fitting additional Gaussian profiles with the same width as the spectrometer resolution. While we detect some excesses of emission lines, none reaches $5\sigma$ significance. 
{The most significant deviation 
%($\chi^2_{\rm without\, line} - \chi^2_{\rm with\, line}=22.4$)
appears at 0.2142 keV, 
corresponding to a $4.3\sigma$ significance. 
Since such a residual is most likely due to our imperfect modeling of the strong Mg X emission line at 0.2142 keV, 
we set the conservative constraint.} 
In Fig.~\ref{Fig:Cons}, we project the 95\% C.L. upper limits from Chandra and eROSITA onto the ($m_{\rm a}$, $g_{a\gamma \gamma}$) and ($g_{aee}$, $g_{a\gamma \gamma}$) planes, shown in the left and right panels, across the entire observational frequency range.

In the left panel of Fig.~\ref{Fig:Cons}, we set the ALP-electron coupling to $g_{aee}=10^{-11}$ (where Compton and bremsstrahlung production dominate) and $g_{aee}\leq10^{-15}$(where photon coalescence production is overwhelming).
We find a stringent upper limit on the coupling $g_{a \gamma \gamma}$ in the ALP mass range between $0.16$ to $4$ keV corresponding to Chandra measurement, and $0.4$ to $16$ keV corresponding to eRosita measurement. 
%\mkred{After accounting for the uncertainties in the stellar models, our $g_{a \gamma \gamma}$ limits exhibit minor variations within $10\%$.}
%\sout{While the left panel uses two benchmark values,} 
The right panel explores the relationship between $g_{aee}$ and $g_{a \gamma \gamma}$ as a complement. 
In the right panel of Fig.~\ref{Fig:Cons}, we present the 95\% limits on the correlation between $g_{aee}$ and $g_{a\gamma\gamma}$ for different mass $m_a$. 
The ALP masses of {4 and 8 keV} for eROSITA and 1 and 4 keV for Chandra reflect different energy ranges of two detectors. 
All four production channels depend on $g_{a \gamma \gamma}$, while the Compton and bremsstrahlung process also depends on $g_{aee}$. 
With small $g_{aee}$ values (cyan region), the Compton and bremsstrahlung process are suppressed, leading to a plateau in the constraint on $g_{a\gamma\gamma}$, corresponding to the limits of $g_{aee}\leq10^{-15}$ in the left panel of Fig.~\ref{Fig:Cons}.
On the contrary, as the value of $g_{aee}$ increases, the Compton and bremsstrahlung process becomes more pronounced, resulting in enhanced constraints on $g_{a\gamma\gamma}$. For specific mass of ALP, the enhanced X-ray limits become sensitive to the product $g_{aee} g_{a\gamma\gamma}$, which is displayed in the right panel of Fig.~\ref{Fig:Cons}.

Our limits on $g_{a\gamma\gamma}$ extend to ALP masses as low as 0.16 keV, benefiting from the lower X-ray energy threshold of Chandra and eROSITA. 
In the absence of the ALP-electron coupling ($g_{aee}\leq10^{-15}$) for keV-scale ALP, our constraints on $g_{a\gamma\gamma}$ are about one to two orders of magnitude stronger than the solar X-ray limits reported in Ref. \cite{Beaufort:2023zuj} for the mass below 4 keV. Once a strong axion-electron coupling ($g_{aee}\sim 10^{-11}$) is introduced, the $g_{a\gamma\gamma}$ (or corresponding $g_{a\gamma\gamma}  g_{aee}$) limits can be improved by up to three orders of magnitude.

This significant improvement results from three important factors: 
\begin{itemize}
\item The solar X-ray emission below $\sim 2$ keV is too bright to be directly recorded by the modern sensitive detectors~\cite{Weisskopf:2000tx}. While for  Alpha-Centauri, thanks to its greater distance, such X-rays have been accurately measured by Chandra and eROSITA.
%provide better X-ray data below $2$ keV from Alpha-Centauri} compared to the %SphinX and NuSTAR from the Sun~\cite{Weisskopf:2000tx, Beaufort:2023zuj};
\item  While Ref.~\cite{Beaufort:2023zuj} conservatively assumes ALP-induced X-ray fluxes are below the total observed solar fluxes, our study relies on spectral line analysis while excluding known astrophysical X-ray lines. 
\item We also consider ALPs produced from bremsstrahlung radiation of electrons with ions inside the star. This production channel dominates over Compton process for ALPs with mass below $\sim 4$ keV.
\end{itemize}

In addition to detecting the X-ray spectrum from gravitationally trapped ALPs, we compare our constraints with those from other observations. 
The left panel of Fig.~\ref{Fig:Cons} shows limits from HB stars in globular clusters (cyan)~\cite{Carenza:2020zil}, Type II supernova SN 2023ixf (grey)~\cite{Muller:2023pip}, and the Neutron Star merger GW170817 (purple)~\cite{Dev:2023hax}. Our $g_{a\gamma\gamma}$ limits from Alpha Centauri are improved 
%on these 
by about two orders of magnitude for ALP masses around $2$ keV. Finally, future X-ray telescopes (e.g., XRISM~\cite{2024arXiv240619910S}, AXIS~\cite{Reynolds:2023vvf}, and Athena~\cite{Barcons:2012zb}), with their enhanced capabilities, are expected to significantly advance ALP research.

\begin{acknowledgments}
This work was supported by the National Key Research and Development Program of China (2022YFF0503304, and 2022YFF0503301), the National Natural Science Foundation of China (11921003, and 12003069), the New Cornerstone Science Foundation through the XPLORER PRIZE, the Project for Young Scientists in Basic Research of the Chinese Academy of Sciences (No. YSBR-092), and the Entrepreneurship and Innovation Program of Jiangsu Province. 
This research has made use of data obtained from the Chandra Data Archive provided by the Chandra X-ray Center (CXC). This work is based on data from eROSITA, the soft X-ray instrument aboard SRG, a joint Russian-German science mission supported by the Russian Space Agency (Roskosmos), in the interests of the Russian Academy of Sciences represented by its Space Research Institute (IKI), and the Deutsches Zentrum für Luftund Raumfahrt (DLR). The SRG spacecraft was built by Lavochkin Association (NPOL) and its subcontractors, and is operated by NPOL with support from the Max Planck Institute for Extraterrestrial Physics (MPE). The development and construction of the eROSITA X-ray instrument was led by MPE, with contributions from the Dr. Karl Remeis Observatory Bamberg \& ECAP (FAU Erlangen-Nuernberg), the University of Hamburg Observatory, the Leibniz Institute for Astrophysics Potsdam (AIP), and the Institute for Astronomy and Astrophysics of the University of Tübingen, with the support of DLR and the Max Planck Society. The Argelander Institute for Astronomy of the University of Bonn and the Ludwig Maximilians Universität Munich also participated in the science preparation for eROSITA. 
\end{acknowledgments}

\section*{Supplemental Material}

\subsection{Production of ALPs}
\label{sec:production_of_ALPs}

%An important property of axion-like particles (ALPs) is their coupling to Standard Model particles. 
%Several extensions of the Standard Model predict that axion-like particles (ALPs) are pseudoscalar bosons with an ALP-photon coupling 
%$g_{a \gamma \gamma}$, which allows ALPs to be produced in stellar plasmas through the Primakoff process~\cite{Raffelt:1985nk}. The photon-ALP transition rate for the Primakoff process can be expressed as a phase space integral~\cite{Raffelt:1985nk,DiLella:2000dn}, which is given by

Several extensions of the Standard Model suggest that axion-like particles (ALPs) are pseudoscalar bosons with an ALP-photon coupling $g_{a \gamma \gamma}$. 
This coupling enables ALPs to be produced in stellar plasmas via the Primakoff process~\cite{Raffelt:1985nk}. 
The photon-ALP transition rate for this process can be described as a phase space integral~\cite{Raffelt:1985nk,DiLella:2000dn},
\begin{equation}
\Gamma^{\rm Primakoff}=\frac{g_{a\gamma \gamma}^2 T \kappa^2}{32 \pi^2} \frac{|\textbf{k}|}{\omega} \int d\Omega \frac{|\textbf{k}\times\textbf{p}|^2}{(\textbf{k}-\textbf{p})^2((\textbf{k}-\textbf{p})^2+\kappa^2)},
\label{Gamma_pri_ori}
\end{equation}
where the temperature at production, denoted as $T$, depends on the radius of Alpha Centauri. 
The momenta of the photon and ALP are represented by $\textbf{k}$ and $\textbf{p}$, respectively, while $\omega$ indicates the photon energy. 
The Debye-Hückel screening scale, $\kappa^2$, is defined by 
\begin{equation}
\kappa^2=4\pi\alpha \sum_{i=e,\rm ions} Z_i^2 \frac{n_i}{T}, 
\end{equation}
where $n_i$ is the number density of charged particles. 
In the case of Alpha Centauri, we can only consider electrons, $H^+$, and $He^{2+}$, as the heavier elements can be neglected due to the low metallicity. 
For small momenta, the Primakoff transition rate for ALPs with mass $m_a$ can be expressed as~\cite{Bastero-Gil:2021oky}
\begin{equation}
\Gamma^{\rm Primakoff} \simeq\frac{g_{a\gamma \gamma}^2 T \kappa^2}{32 \pi^2} \left[ \frac{8p^2}{3(\kappa^2+m_a^2)}+\mathcal{O}(p^4) \right]. 
\label{Gamma_pri}
\end{equation}

ALPs with mass around keV can be produced in stellar cores through photon coalescence $\gamma \gamma \rightarrow a$. 
When considering heavier ALPs or stars with higher core temperatures, the photon coalescence process can be more efficient than the Primakoff effect. 
The photon-ALP transition rate for photon coalescence is given by~\cite{Cadamuro:2010cz,Cadamuro:2011fd}
\begin{equation}
\Gamma^{\rm coal.} (T)=\Gamma_{a \gamma \gamma} \frac{m_a^2-4\omega_{\rm p}^2}{m_a^2} \left( \frac{m_a}{E_a} \right) 
\left[ 1+\frac{2T}{p} \ln \frac{1-e^{-(E_a+p)/2T}}{1-e^{-(E_a-p)/2T}} \right],
\label{Gamma_coale}
\end{equation}
where $\omega_{\rm p}=(4 \pi \alpha n_e/m_e)^{1/2}$ is the photon thermal "mass" with electron number density $n_e$ and electron mass $m_e$, and
$\alpha$ is the fine-structure constant. The ALP decay rate into photons can be written as   
\begin{equation}
\Gamma_{a\gamma \gamma}=\frac{g_{a\gamma\gamma}^2 m_a^3}{64\pi}.
\label{Gamma_decay}
\end{equation}
{Notice} the photon coalescence process is suppressed for $m_a \leq 2\omega_{\rm P}$. 
%The lifetime for $m_a$ around keV can be billions of years, allowing them to escape from the interior of Alpha Centauri and effectively accumulate in orbit. 
The ALP lifetime with $m_a$ around keV can be billions of years, enabling it to escape from the interior of Alpha Centauri and accumulate in orbit.

The ALP-electron coupling $g_{aee}$ can be naturally generated in non-hadronic ALP models~\cite{Zhitnitsky:1980tq,Dine:1981rt}, 
allowing ALPs to be produced in the star through the Compton mechanism. 
The Compton transition rate~\cite{Beaufort:2023zuj} is
\begin{equation}
\Gamma^{\rm Comp.} \simeq \alpha g_{aee}^2 n_e \frac{m_a}{m_e^4} \sqrt{m_a^2-\omega_{\rm p}^2}.
\label{Gamma_comp}
\end{equation}
In addition to Compton process, ALPs are also produced by bremsstrahlung radiation of electrons when scattering with ions inside the star, as long as the parameter $g_{aee}$ is not negligible. 
With the same notation and soft ALPs approximation $|\bm k|\equiv \sqrt{\omega^2 - m_a^2} \ll m_a$, the production rate is given from \cite{VanTilburg:2020jvl} as follows
\begin{align}
    \Gamma^{\text{brem.}} \simeq &\frac{\alpha^2 g_{aee}^2\bar {n_{N}}n_e }{32\sqrt{2}\pi^{5/2}m_e^{7/2}T^{3/2}}\nonumber \\
    &\times\int d\epsilon \bigg[\log{\frac{2+2\sqrt{1-\epsilon}-\epsilon + \xi}{\epsilon + \xi}}\times \text{exp}\{-\frac{m_a}{\epsilon T}\}\bigg],\,
\end{align}
where we define the dimensionless parameters $\epsilon \equiv 2m_em_a/{\bm P_e^2}$ and $\xi \equiv \kappa^2/\bm P_e^2$. Notice $\bar{n_i}$ is defined as $\sum_{i}Z_i^2n_i$ for only ions and $\bm P_e$ is the incoming electron momentum.

%%%%%%%%%%%%%%%%%%%%%%%%%%%%%%%%%%%%%%
\subsection{ALPs in orbit}
%%%%%%%%%%%%%%%%%%%%%%%%%%%%%%%%%%%%%%

\begin{figure}
    \centering 
    \includegraphics[width=1.0\linewidth]{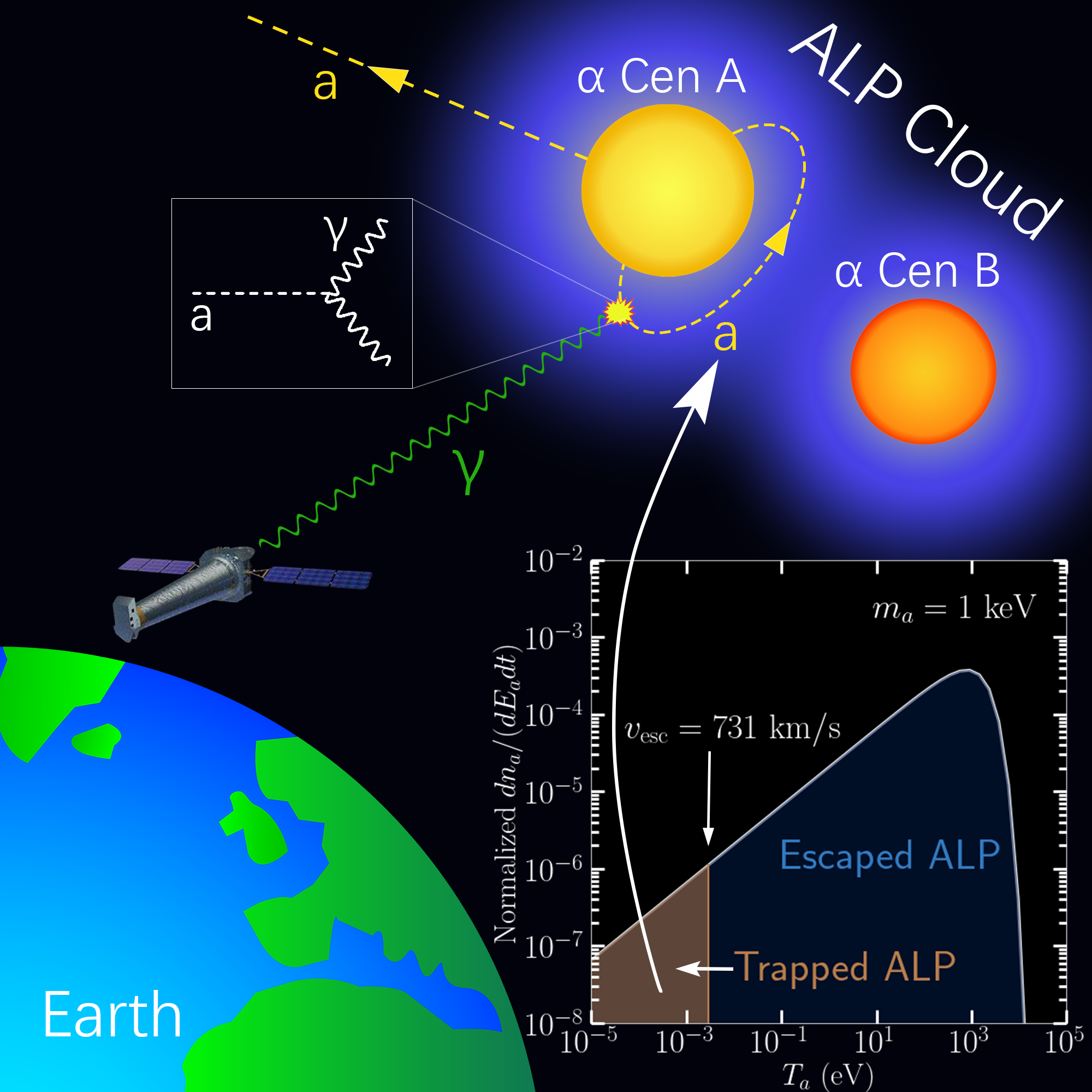}
    \caption{A schematic diagram illustrating stellar ALPs gravitationally trapped by the Alpha Centauri system. For produced ALPs with a mass of $m_a=1$~keV, the number density per unit time and per unit energy as a function of ALP kinetic energy $T_a$ is depicted in the lower right corner. The brown region represents the gravitationally trapped ALPs, whose velocities fall below the escape threshold.}
    \label{fig:schematic} 
\end{figure}

Axion-like particles (ALPs) with masses around the keV scale, produced in Alpha Centauri through processes such as the Primakoff effect, photon coalescence, and Compton scattering, are sufficiently non-relativistic to become gravitationally trapped by the system. Once trapped, these ALPs can orbit Alpha Centauri and gradually accumulate over time. This accumulation process is tied to the stellar evolution of Alpha Centauri and may continue for billions of years, depending on the lifetime of the ALPs, potentially leading to a significant increase in their population within the system, as schematically illustrated in Fig. \ref{fig:schematic}. Following the methods outlined in Ref.~\cite{Bastero-Gil:2021oky}.

The number density of the total ALPs production $n_a$ is derived from the Boltzmann equation 
\begin{equation}
\frac{dn_a}{dt}\simeq\frac{1}{(2\pi)^3}\int d^3p
\Gamma_{\rm Prod.} f_a^{\rm eq}(E_a), 
\label{dnadt}
\end{equation}
where $\Gamma_{\rm Prod.}$ is the production rate of ALPs in Alpha Centauri and $f_a^{\rm eq}(E_a)= (e^{E_a/T}-1)^{-1}$ represents the equilibrium distribution function of ALPs, following a Bose-Einstein distribution.

If the orbit of an ALP is bounded, the radial velocity becomes zero at the apohelion, corresponding to a distance $r$ from the center of the star. We impose the relation between the initial phase space of the produced ALP and apohelion distance by energy conservation. This leads to the condition: 
\begin{equation}
    h(\bm v_0; r,r_0) \equiv v_{r,0}^2 + v_{\phi,0}^2\left[1 - \left(\frac{r_0}{r}\right)^2 \right] - 2\left[\Phi(r_0) - \frac{GM}{r} \right] =0,\,
    \label{Bounded_trajectories}
\end{equation}
where $v_r$ and $v_{\phi,0}$ represent the radial velocity and angular velocity of the ALPs, respectively. The subscript “$0$” refers to the initial condition of ALPs for radius and velocities for produced ALPs. The gravitational potential $\Phi(r_0)$ within the star is for main sequence phase and can be computed using the density profile of Alpha Centauri, derived from the \texttt{MESA}~\cite{Paxton:2010ji,Paxton:2013pj}.

We calculate the number density of ALPs at a distance $r$ from the Alpha Centauri.  For ALPs produced within the star at a distance 
$r_0$, they must satisfy the constraint in Eq.~\eqref{Bounded_trajectories}. By integrating over the entire phase space and the volume of the star, the number density of trapped ALPs is given by:
\begin{equation}
    \frac{dn_a^{\rm trap}}{dt} = \left(\frac{m}{2\pi}\right)^3\int^{R_s}_0 dr_0 ~4\pi r_0^2 \int d^3v_0 N\delta \left( h(\bm v_0) \right)\Gamma_{\rm Prod.}f^{\rm{eq}}(E_a),\,
    \label{eq:trapped_density}
\end{equation}
%\begin{equation}
%v_{r,0}^2+r_0^2 \Dot{\phi_0^2} (1-\bar{r}_0^2 \bar{u}_r^2)=2\frac{GM}{R_{s}}(\bar{\Phi}_{\rm G}(\bar{r}_0)-\bar{u}_r), 
%\label{Bounded_trajectories}
%\end{equation}
%where $v_r$ and $\Dot{\phi}$ represent the radial velocity and angular velocity of the ALPs, respectively. The subscript “$0$” refers to the initial condition of ALPs for radius and velocities inside Alpha Centauri and $\bar{u}_r=R_s/r$. 
%The gravitational potential $\Phi_{\rm G}$ inside Alpha Centauri is given by
%\begin{equation}
%\begin{split}
%\Phi_{\rm G}(r<R_{s})&=-4\pi G \left( \frac{1}{r} \int_0^r r'^2 \rho(r')d r' +\int_r^{R_s} r' \rho(r')dr' \right)\\
%&=-\frac{GM}{R_{s}}\bar{\Phi}_{\rm G}(\bar{r}),
%\label{potential}
%\end{split}
%\end{equation}
where $m$ and $R_s$ are the mass and radius of the star and $\delta(x)$ is Dirac delta function.  The factor $N$ normalizes the probability of finding a particle with given energy at radius $r$, such that 
\begin{equation}
    \int d^3r~ N\delta \left(h(\bm v_0; r,r_0)\right ) = 1.\,
    \label{eq:nomralization}
\end{equation}
%The mass density of Alpha Centauri is represented by $\rho$. 
%The overline "bar" denotes normalization to the Alpha Centauri value. After performing the momentum integration on Eq.~\eqref{dnadt}, according to Ref.~\cite{Bastero-Gil:2021oky}, the number density per unit time of trapped ALPs can be written as:
After solving for $N$ in Eq.~\eqref{eq:nomralization} and substituting it back into Eq.~\eqref{eq:trapped_density} with condition Eq.~\eqref{Bounded_trajectories}, 
the number density of trapped ALPs per unit time can be expressed as 
\begin{align}
    \frac{dn_a^{\rm trap}}{dt} = &\frac{m_a^3R_s}{4\pi^3 r^4}\int_0^{R_s}~dr_0
    r_0^2
    \Gamma_{\rm Prod.}f_a^{eq}(E_a)%\nonumber\\
    %& 
     \sqrt{2\left(\Phi(r_0)- \frac{GM}{r}\right)}.\,
\end{align}

%\begin{equation}
%\frac{dn_a^{\rm trap}}{dt}=\frac{m_a^3}{4\pi^3 \bar{r}^4} \int_0^1 \bar{r}_0^2 d\bar{r}_0 \sqrt{B(\bar{r},\bar{r}_0)} \Gamma_{\rm Prod.}(T) f_a^{\rm eq}(E), 
%\label{trapped_ALPs}
%\end{equation}
%where $B(\bar{r},\bar{r}_0)=2(\bar{\Phi}_{\rm G}(\bar{r}_0)-\bar{u}_r)$ is defined, and $E \simeq m_a(1+\frac{GM}{2R_s}\sqrt{B(\bar{r},\bar{r}_0)})$ is the maximum energy that ALPs can be gravitationally bound by Alpha Centauri.

Moreover, ALPs bound in orbit slowly decay into photons, $a \rightarrow \gamma \gamma$ , producing X-rays that can be detected on Earth. The evolution of the number density of ALPs in orbit $n_a^{\rm orb}$ is governed by:
\begin{equation}
\Dot{n}_a^{\rm orb}(t)=\Dot{n}_a^{\rm trap}-\Gamma_{a \gamma \gamma} n_a^{\rm orb}(t),
\label{ALPs_in_orbit_ori}
\end{equation}
where $t$ is the accumulated time of the ALPs in orbit. 
Finally, the number density of ALPs in orbit can be written as  
\begin{equation}
n_a^{\rm orb}(t)=\frac{1-\exp\left(-\Gamma_{a \gamma \gamma}t\right)}{\Gamma_{a \gamma \gamma}}
\frac{dn_a^{\rm trap}}{dt}.
\label{ALPs_in_orbit}
\end{equation}

%%%%%%%%%%%%%%%%%%%%%%%%%%%%%%%%%%%%%%
\subsection{The best-fit energy spectra}

In Fig.~\ref{fig:spec} of the main text, the Chandra spectrum was rebinned by a factor of 5 in each energy bin relative to the original data. However, the original data are used in the fitting and analysis processes. The data points are plotted in black with error bars. 
In the modeling of the spectra, we still use the raw unbin data for spectral background fitting and emission line search. We perform adaptive rebinning of the X-ray photon counts from Chandra (eROSITA) X-ray satellite by successively grouping energy bins until a signal-to-noise ratio of 100 (7) is achieved. To prevent excessive bin width, groupings are capped at a maximum of 100 (7) original bins per final bin. The rebin here is just for the convenience of seeing the background model when drawing.

\begin{table}
  \centering
  \renewcommand{\arraystretch}{1.2}
  \setlength{\tabcolsep}{6pt}
  \caption{Background Models of Chandra and eROSITA spectrum}
  \label{tab:bkgfit}
  \begin{threeparttable}
  \begin{tabular}{l|c|c}
    \hline\hline
       & Chandra  & eROSITA  \\
    \hline\hline
    $T_{\rm blackbody}$ (keV) & $0.042 \pm 0.001$  & $0.132 \pm 0.005$  \\
    $N_{\rm blackbody}/10^{-5}$ & $5.84 \pm 0.12$  & $8.81 \pm 0.70$  \\
    \hline
    $T_{\rm Apec}$ (keV) & $0.139 \pm 0.001$  & $0.061 \pm 0.003$  \\
    $N_{\rm Apec}/10^{-4}$ & $5.84 \pm 0.12$  & $850.1 \pm 148.0$  \\
    \hline\hline
  \end{tabular}
  \end{threeparttable}
\end{table}

%The background models for both spectra are assumed to originate from stellar soft X-ray radiation. As detailed in the main text, we analyzed the Chandra/LETG spectrum in the range between $0.08$ and $2.0$ keV using the BlackBody and Apec models in \texttt{Xspec} as the background model. For the eROSITA spectrum, we similarly employed BlackBody and APEC models in the range $0.2<E_\gamma/{\rm keV}<8.0$ through \texttt{Xspec} to model the stellar background.The spectral fitting results indicate plasma temperatures of 0.13 keV and 0.07 keV for the Chandra and eROSITA observations, respectively.Table~\ref{tab:bkgfit} shows the best-fit parameters of the Apec and BlackBody models.
The background models for both spectra are assumed to originate from stellar soft X-ray radiation. 
As described in the main text, we analyzed the Chandra/LETG spectrum in the energy range from 0.08 to 2.0 keV using BlackBody and Apec models in \texttt{Xspec} for the background. 
Similarly, for the eROSITA spectrum (ranging from 0.2 to 8.0 keV), we applied BlackBody and Apec models in \texttt{Xspec} to model the stellar background. Since the number of X-ray photons is relatively small, we used the Cash statistical method for fitting \cite{1979ApJ...228..939C}.
The spectral fitting results yield plasma temperatures of 0.14 keV for Chandra and 0.06 keV for eROSITA. 
The best-fit parameters of the Apec and BlackBody models used in \texttt{Xspec} code are shown in Table~\ref{tab:bkgfit}.
The parameters $T_{\rm blackbody}$ and $T_{\rm Apec}$ represent the temperatures of the blackbody and plasma, respectively, 
while $N_{\rm blackbody}$ and $N_{\rm Apec}$ are their corresponding distribution normalizations.

%%%%%%%%%%%%%%%%%%%%%%%%%%%%%%%%%%%%%%
\subsection{Discussion of astrophysical uncertainties}
%%%%%%%%%%%%%%%%%%%%%%%%%%%%%%%%%%%%%%

\begin{table}[!ht]
  \centering
  \renewcommand{\arraystretch}{1.2}
  \setlength{\tabcolsep}{6pt}
  \caption{Details of $\alpha$ Centauri A and B stellar parameters\tnote{$\star$}}
  \label{tab:alpha_centauri}
  \begin{threeparttable}
  \begin{tabular}{l|c|c}
    \hline\hline
    %\multicolumn{2}{c}{$\boldsymbol{\alpha}$ \textbf{Centauri A}} & \multicolumn{1}{c}{$\boldsymbol{\alpha}$ \textbf{Centauri B}} \\
       & $\boldsymbol{\alpha}$ \textbf{Centauri A}  & $\boldsymbol{\alpha}$ \textbf{Centauri B}  \\
    \hline\hline
    Mass\tnote{$\divideontimes$}~ ($M_{\odot}$) & $1.0788 \pm 0.0029$  & $0.9092 \pm 0.0025$  \\
    %Radius ($R_{\odot}$) & $1.2175 \pm 0.0055$  & $0.8591 \pm 0.0036$  \\
    %Luminosity & $1.5059 \pm 0.0019$ & $L_{\odot}$ & $0.4981 \pm 0.0007$ & $L_{\odot}$ \\
    %Surface gravity (log $g$) & $4.30$ & cgs & $4.37$ & cgs \\
    %Temperature & $5,804 \pm 13$ & K & $5,207 \pm 12$ & K \\
    Metallicity [z]\tnote{$\dag$} & $0.027 \pm 0.005$  & $0.027 \pm 0.005$ \\
    %Rotation & $28.3 \pm 0.5$ & d & $36.7 \pm 0.3$ & d \\
    %Rotational velocity ($v \sin i$) & $2.7 \pm 0.7$ & km/s & $1.1 \pm 0.8$ & km/s \\
    Age\tnote{$\dag$} (Gyr) & $5.26 \pm 0.95$  & $5.26 \pm 0.95$  \\
    \hline\hline
  \end{tabular}
  \begin{tablenotes}
        \footnotesize
        %\item[$\star$] \url{https://en.wikipedia.org/wiki/Alpha_Centauri}
        \item[$\divideontimes$] The mass of the Alpha Centauri binary system is from Ref.~\cite{2021AJ....162...14A}
        \item[$\dag$] The age and metallicity of the Alpha Centauri binary system is from Ref.~\cite{2018ApJ...864...99J}
      \end{tablenotes}
  \end{threeparttable}
\end{table}

Astrophysical uncertainties in ALP production arise from stellar parameters: mass, metallicity, and age. 
Larger stellar masses raise core temperatures, while higher metallicity enhances opacity and radiation absorption, both affecting ALP production. 
Older stars allow trapped ALPs to accumulate longer in orbit, increasing their number density. 
Table~\ref{tab:alpha_centauri} summarizes these parameters (with uncertainties) for Alpha Centauri A and B, based on~\cite{2021AJ....162...14A,2018ApJ...864...99J}.

\begin{figure*}[htbp]
%\centering 
%\includegraphics[width=0.49\linewidth]{reply_to_referee/diff_single.pdf}
\includegraphics[width=0.49\linewidth]{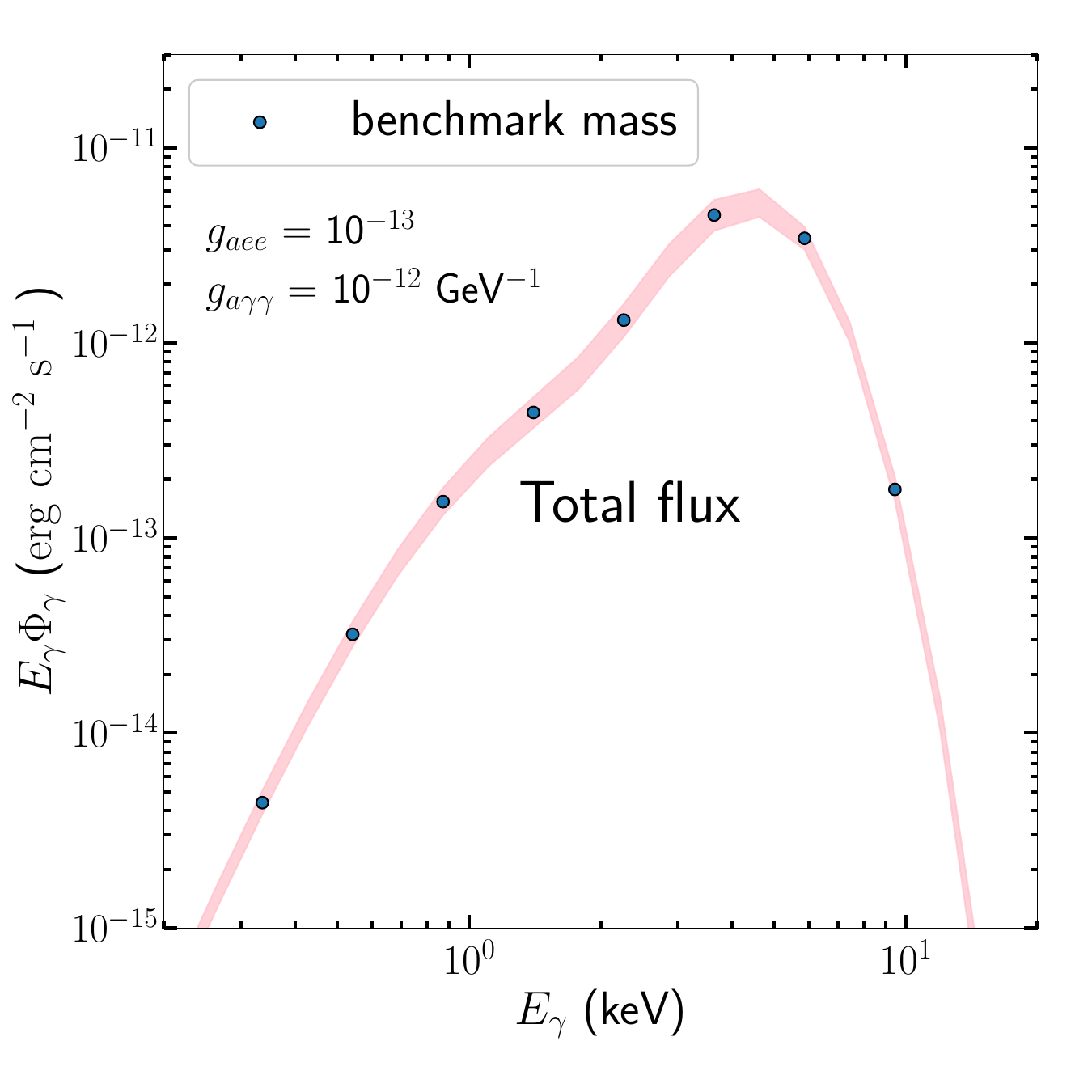}
\includegraphics[width=0.49\linewidth]{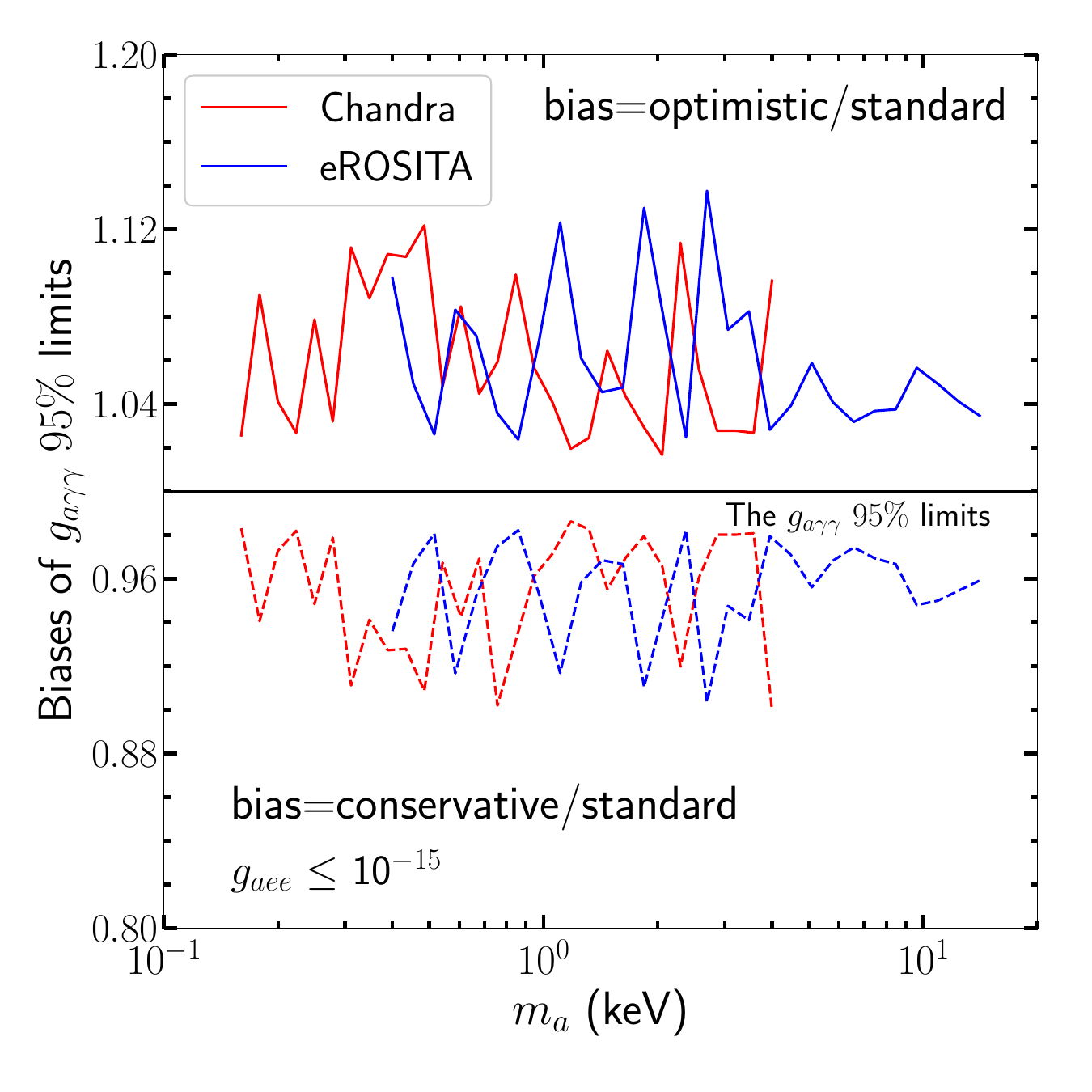}
\caption{\textbf{Left panel:} Photon flux induced by ALPs as a function of photon energy. 
The blue points represent theoretical predictions for different ALP masses, $m_a = 2 E_\gamma$, 
while the pink region reflects the range of astrophysical models and parameter uncertainties. 
\textbf{Right panel:} Bias of the 95\% upper limits. 
The solid black line shows the Chandra and eROSITA upper limits with $g_{aee} \leq 10^{-15}$ from Fig.~\ref{Fig:Cons} of the main text. 
The maximum (solid lines) and minimum (dashed lines) bias, due to astrophysical models and uncertainties, are shown for Chandra (red) and eROSITA (blue).
}
\label{fig:diff_flux} 
\end{figure*}

In addition to stellar parameters, we also examine stellar convection models and element diffusion~\cite{2018ApJ...864...99J} by using \texttt{MESA}~\cite{Paxton:2010ji}. 
For convection, we vary the relevant assumptions: 
\begin{itemize}
    \item[(i)] %\textbf{Mixing length}:\\
    {\it The mixing length}, determined by the parameter $\alpha$ and local pressure scale height $H_p$, reflects the efficiency of convective energy transport.
A larger mixing length means convective elements travel farther, increasing convection efficiency and steepening temperature gradients.  For Alpha Centauri A and B, we adopted $\alpha_{A}=1.7\pm0.3$ and $\alpha_{B}=2\pm0.4$, respectively~\cite{2018ApJ...864...99J}. 
    
    \item[(ii)] %\textbf{The Schwarzschild and Ledoux criteria}:\\
{\it The Schwarzschild criterion and the Ledoux criterion} are two conditions that determine the boundary of convection within a star~\cite{Paxton:2010ji}. 
    The Schwarzschild criterion applies when there is no composition gradient, while the Ledoux criterion generalizes the Schwarzschild criterion by taking into account composition gradients. 
    In the \texttt{MESA} code, the Schwarzschild criterion is used by default, and we also study the impact of the Ledoux criterion on ALP production.
    
    \item[(iii)] %\textbf{Convective overshooting}:\\
{\it Convective overshooting} occurs when convective elements, due to inertial effects, exceed the convection zone boundaries and penetrate into the radiative zone, potentially enlarging the convective core and hydrogen-burning region. We considered overshooting parameters similar to the Sun as well as scenarios with no overshooting.
\end{itemize}

For elemental diffusion, it affects opacity, nuclear reaction rates, and the chemical composition distribution. 
In \texttt{MESA} code, elemental diffusion is turned off by default. 
We calculate the differences in ALP production between scenarios with heavy diffusion and without diffusion.

We vary the parameters and models discussed earlier to calculate the maximum and minimum ALP-induced photon flux, 
representing our optimistic and conservative results in comparison to the standard settings used in this work.
The differences between these results are shown in Fig.~\ref{fig:diff_flux}. 
The left panel illustrates the ALP-induced photon flux for both optimistic and conservative settings (pink region), with ALP-electron coupling $g_{aee} = 10^{-13}$ and ALP-photon coupling $g_{a\gamma\gamma} = 10^{-12}~{\rm GeV}^{-1}$, as in Fig.~\ref{fig:spec} of the main text. 
The blue points represent theoretical predictions for various ALP masses, $m_a = 2 E_\gamma$. 
We observe a maximum $55\%$ difference between the optimistic and conservative results within the photon energy range of $0.08~{\rm keV} < E_\gamma < 8~{\rm keV}$, 
leading to a bias of $\pm 14\%$ for the $g_{a\gamma\gamma}$ 95\% upper limits in Fig.~\ref{Fig:Cons} of the main text. 
Thank to the precise measurements of the mass and metallicity of the Alpha Centauri binary system~\cite{2021AJ....162...14A,2018ApJ...864...99J}, the bias in our results is very small.

\bibliographystyle{apsrev}
%\bibliography{references}

\end{document}